\documentclass[12pt]{iopart}%
\usepackage{iopams}
\usepackage{graphicx}
\usepackage{amssymb}
\usepackage{amsfonts}%
\providecommand{\U}[1]{\protect\rule{.1in}{.1in}}
\newtheorem{theorem}{Theorem}

\newtheorem{remark}[theorem]{Remark}
\begin{document}

\title{Stable and fast semi-implicit integration of the stochastic Landau-Lifshitz equation}
\author{J.H. Mentink$^{1}$, M.V. Tretyakov$^{2}$, A. Fasolino$^{1}$,
M.I.~Katsnelson$^{1}$, Th. Rasing$^{1}$}

\begin{abstract}
We propose new semi-implicit numerical methods for the integration of the
stochastic Landau-Lifshitz equation with built-in angular momentum
conservation. The performance of the proposed integrators is tested on the 1D
Heisenberg chain. For this system, our schemes show better stability
properties and allow us to use considerably larger time steps than standard
explicit methods. At the same time, these semi-implicit schemes are also of
comparable accuracy to and computationally much cheaper than the standard
midpoint implicit method. The results are of key importance for atomistic spin
dynamics simulations and the study of spin dynamics beyond the macro spin approximation.
\end{abstract}

\address{$^1$Radboud University Nijmegen, Institute for Molecules and Materials,
Heijendaalseweg 135, 6525 AJ, Nijmegen, The Netherlands.} \ead{J.Mentink@science.ru.nl}

\address{$^2$Department of Mathematics, University of Leicester,
Leicester, LE1 7RH, United Kingdom.} \ead{M.Tretyakov@le.ac.uk}



\section{Introduction}

Dynamics of magnetic materials have been theoretically studied for many years
starting from the seminal work by Landau and Lifshitz \cite{landau1935} (see,
e.g., the monographs \cite{akhiezer1968,vonsovsky1974,aharoni2000}). The
current interest in this area is rapidly growing due to new important fields
of applications such as spintronics \cite{zutic2004} and laser-induced
ultrafast spin dynamics \cite{gerrits2002,kimel2005,koopmans2005,melnikov2008}%
. In many situations such as the interaction of domain walls with pinning
centers \cite{novoselov2003}, there are atomic-scale inhomogeneities which
require multiscale simulations bridging macroscopic and microscopic lengths
\cite{dobrovitski2003}. In \cite{antropov1995,antropov1996} a method of
\textit{ab initio} spin dynamics was suggested relating first-principle
electronic structure calculations with Landau-Lifshitz-type dynamics of
classical spins within the framework of the rigid-spin approximation.

Thus, Atomistic Spin Dynamics (ASD) simulations are important from many points
of view. To do calculations at finite temperatures, there are two main
approaches: a generalized Nose-Hoover (Bulgac-Kuznecov) thermal bath or the
Langevin (stochastic) dynamics \cite{antropov1996}. The first method has
fictitious dynamics, and hence it can be used to simulate equilibrium
properties only. The Langevin spin dynamics with first-principle magnetic
interaction parameters has recently been implemented \cite{skubic2008} and
applied for simulating dilute magnetic semiconductors \cite{hellsvik2008} and
spin glasses \cite{skubic2009}. Langevin spin dynamics are also used as a
phenomenological simulation tool, not connected with first-principle
theory. An implementation of this type was reported in \cite{nowak2001} and
applied to laser-induced magnetization dynamics \cite{kazantseva2008}.

The heart of Langevin spin dynamics simulations is integration of the
stochastic Landau-Lifshitz (SLL) equation for each atomic spin. This equation
is non-linear and analytical solutions for interacting systems exist for two
spins only. In systems of interest for applications the number $n$ of spins is
typically of order $10^{6}$ and the integration should be done numerically.
Due to the interactions, one has to solve a system of $3n$ coupled non-linear
equations. To compute quantities in equilibrium, this very large system should
be simulated over long time intervals, usually from $10$ fs to $1$ ns. This is
a challenging computational task.

Thus, ASD requires effective numerical integrators for the SLL equation. Due
to the large system size and long simulation time, such numerical methods
should be, on the one hand, sufficiently stable and on the other hand very
fast. The latter rules out the use of fully implicit integrators such as the
implicit midpoint (IMP) scheme (see its application for Langevin spin
dynamics, e.g., in \cite{daquino2006}). Despite its superior stability
properties which allows large step sizes, typically 10 fs, IMP is slow in
practice since the implicitness requires solution of $3n$ non-linear coupled
equations at every time step. Langevin spin dynamics simulations have often
been based on the Heun method \cite{skubic2008,nowak2001}, which has the
advantage of being fast in terms of the number of operations per time step.
However, this method has poor stability properties requiring a relatively
small step size, typically ranging from 0.01 fs to 1 fs, depending on the
implementation. We also note that since the accuracy of the first-principle
magnetic interaction parameters is limited to $10\%$, the accuracy of
numerical methods is, to some extent, less important here than their stability
(in the sense of the ability to use larger step sizes for long time
simulations). Hence, both the standard implicit and explicit numerical
integrators are not optimal for ASD and it is desirable to develop a numerical
method that is both stable and fast. Also, ASD simulations are often used to
study systems with different interactions and/or different symmetries.
Therefore, in addition we should require from numerical integrators for ASD to
be universal in their implementation. Such a method is proposed in this paper.

As is known from the deterministic (\cite{HLW02,frank1997,arponen2004} and the
references therein) and stochastic \cite{MRT02b,MT,RRM09} numerical
approaches, to numerically integrate dynamical systems over long time
intervals with relatively large step sizes, it is advisable to preserve
geometrical properties of the continuous dynamics. Therefore, one should
construct and use geometrical integrators for ASD.

In the case of the deterministic Landau-Lifshitz (LL) equation, there are
geometric integrators \cite{frank1997,arponen2004,krech1998,steinigeweg2006}
that are both stable and fast. Usually, these schemes are semi-implicit.
Unlike IMP, a semi-implicit method requires only the solution of $3$ linear
coupled equations for each spin individually. However, implementation of these
methods depends on symmetry and interactions in a system under consideration,
which makes it difficult to use them for models with arbitrary lattice structures.

Further, semi-implicit methods for the deterministic LL equation are also
considered in the review \cite{banas2005}. Being based on IMP, they have the
potential to combine stability and low computational costs like the geometric
integrators but with the advantage of a universal implementation. In this
paper we use the idea of semi-implicitness to derive new numerical methods for
Langevin spin dynamics simulations, which are both stable and fast and allow
universal implementation. In particular, we show that, due to the enhanced
stability, our semi-implicit integrator (named SIB) allows time steps by a
factor of $10\div10^{3}$ larger than the standard Heun method.

This paper is organized as follows. In Section~\ref{sec_model}, we formulate
the problem in mathematical terms, introduce the necessary notation and
examine the conservation properties of the SLL equation. In
Section~\ref{sec_numermethod}, we propose two new semi-implicit methods (SIA
and SIB) and recall the Heun scheme and IMP. Both SIA and SIB intrinsically
preserve the length of individual spins while SIB (like IMP) also possesses other
conservation properties in the deterministic case. The later is apparently the
reason for the superiority of SIB which is the numerical method of our choice
for ASD. In Section~\ref{sec_exp}, we present some results of numerical
experiments. We first test the considered numerical methods in the
deterministic case without damping, using a simple system of two interacting
spins. Then the 1D Heisenberg chain is used as a test system for the
stochastic case. In the last section, we draw conclusions and recommendations
for future work. Two appendices are included to provide some auxiliary
knowledge of stochastic numerics and about ergodicity of the SLL equation.

\section{Mathematical model\label{sec_model}}

In this section we formulate the problem in mathematical terms and introduce
the necessary notation. In addition, we discuss why we use the Stratonovich
interpretation for the stochastic LL equation. Finally, we examine the
properties of the solution of the equations under study.

The (deterministic) Landau-Lifshitz equation in dimensionless variables can be
written in the form:
\begin{equation}
\frac{dX^{i}}{dt}=-X^{i}\times B^{i}(\mathbf{X})-\alpha X^{i}\times\lbrack
X^{i}\times B^{i}(\mathbf{X})]\ ,\ \ \ i=1,\ldots,n,\label{lln}%
\end{equation}
where $n$ is the number of spins, $X^{i}=(X_{x}^{i},X_{y}^{i},X_{z}^{i}%
)^{\top}$ are three-dimensional column-vectors representing unit spin
vectors\footnote{In the paper, we follow the standard notation of the theory
of stochastic differential equations and use capital letters to denote
solutions of differential equations while we use small letters for the initial
data and for corresponding \textquotedblleft dummy\textquotedblright%
\ variables.} and $\mathbf{X}=(X^{1^{\top}},\ldots,X^{n^{\top}})^{\top}$ is a
$3n$-dimensional column-vector formed by the $X^{i}$; $B^{i}$ is the effective
field acting on spin $i;$ $\alpha\geq0$ is the damping parameter. In
(\ref{lln}) the time is normalized by the precession frequency $\omega
_{\hat{B}}=\gamma\hat{B}$, where $\hat{B}$ is some reference magnetic field
strength, and the effective field $B=(B^{1^{\top}},\ldots,B^{n^{\top}})^{\top
}$ is also normalized by $\hat{B}$ and is given by
\begin{equation}
B(\mathbf{\mathbf{x}})=-\nabla H(\mathbf{x}),\label{ham}%
\end{equation}
where $H$ is the Hamiltonian of the problem. Then
\[
B^{i}(\mathbf{x})=-\nabla_{i}H(\mathbf{x}),
\]
where $\nabla_{i}$ is the gradient with respect to the Cartesian components of
the effective magnetic field acting on spin $i$.

For atomistic spin dynamics, the most important contributions to the
Hamiltonian are the Heisenberg exchange for the interaction between the spins
$H_{\mathrm{ex}}$, the Zeeman energy for the interaction with an external
field $H_{\mathrm{ext}},$ and the uniaxial anisotropy $H_{\mathrm{ani}}$
defining a preferential direction of the spins. Therefore we consider here the
following Hamiltonian for our problem:
\begin{equation}
H=H_{\mathrm{ex}}+H_{\mathrm{ext}}+H_{\mathrm{ani}},\label{Ham2}%
\end{equation}
where
\begin{eqnarray*}
H_{\mathrm{ex}}(\mathbf{x})  & =-\sum_{i\neq j}J_{ij}{x}^{i}{x}^{j}%
,\ \ \ \ H_{\mathrm{ext}}(\mathbf{x})=-B_{0}\sum_{i}{x}^{i},\\
H_{\mathrm{ani}}(\mathbf{x})  & =K\sum_{i}({x}^{i}{e}_{K})^{2}.
\end{eqnarray*}
Here $J_{ij}$ are the exchange parameters, $B_{0}$ is the uniform external
field, $K$ is the strength of the anisotropy, and ${e}_{K}$ is a unit vector
that defines the anisotropy axis. Note that with these contributions to the
Hamiltonian the effective fields $B^{i}$ are linear in $x$. In realistic
materials usually $|J_{ij}|\gg|{B_{0}}|\gg|K|$. For the exchange parameters
themselves, typically $J_{i(i+1)}\gg J_{i(i+j)}$, $j>1$, i.e., all spins
interact with each other but the nearest-neighbor interactions dominate. Since
all the spins interact, Eq.~(\ref{lln}) involves simultaneous solution of a
$3n$ system of non-linear equations. Due to the interactions between the
spins, each effective field $B^{i}$ is time-dependent and Eq.~(\ref{lln}) has
in general no analytical solution. As a result, efficient numerical methods
are required to study spin systems. In turn, the time-dependence of the
effective field is usually considered as the main source of instability in the
numerical integration.

In order to perform spin dynamics at finite temperature, fluctuations are
included according to the Brownian motion approach for spins by adding
fluctuating torques to Eq.~(\ref{lln}) \cite{kubo1970,brown1963}. The
stochastic Landau-Lifshitz (SLL) equation is then given by
\begin{eqnarray}
\frac{dX^{i}}{dt}=-X^{i}\times(B^{i}(\mathbf{X})+b^{i})-\alpha X^{i}%
\times\lbrack X^{i}\times(B^{i}(\mathbf{X)}+b^{i})]\ ,\label{sll}\\
\ i=1,\ldots,n,\ \nonumber
\end{eqnarray}
where the fluctuating magnetic fields $b^{i}$ are uncorrelated Gaussian white
noises interpreted in the sense of Stratonovich and
\begin{equation}
\left\langle b_{l}^{i}(t)\right\rangle =0,\ \ \left\langle b_{l}^{i}%
(t)b_{k}^{j}(0)\right\rangle =2D\delta_{ij}\delta_{lk}\delta
(t)\ ,\ \ i=1,\ldots,n, \label{white}%
\end{equation}
with $\left\langle \cdot\right\rangle $ denoting ensemble averages and $l,$
$k=x,y,z$ labeling the Cartesian coordinates while $D$ is the strength of the
fluctuations. According to the fluctuation dissipation theorem, we choose
\begin{equation}
D=\frac{\alpha}{(1+\alpha^{2})}\frac{k_{b}T}{\hat{X}\hat{B}}\ , \label{dd}%
\end{equation}
where $\hat{X}$ is the (non-normalized) magnetization of each spin.

Note that (\ref{sll}) is a differential equation with multiplicative noise
which requires from us to specify in which sense we interpret the stochastic
equation \cite{Gard}. As said above, we use here the Stratonovich
interpretation following \cite{kubo1970}. This choice can be motivated as
follows. First of all, the Stratonovich interpretation (contrary to any other
one and, in particular, to the Ito interpretation) leads to preservation of
the individual spin length (see (\ref{spin_pres}) below) by (\ref{sll}), which
is very important to model spin systems (see also a similar discussion in \cite{daquino2006}). Further, it is natural to
model a perturbation of the Landau-Lifshitz dynamics by Gaussian noise with a
finite bandwidth spectrum (i.e., by a colored noise \cite{Gard}), possibly
with a very short correlation time. The white noise $b(t)$ in (\ref{sll}) has
zero correlation radius (see (\ref{white})) and a spectrum with infinite
bandwidth. This noise is a convenient idealization which can be viewed as an
approximation of the colored noise with short correlation time. Indeed, if we
consider a sequence of solutions $X_{n}(t)$ of the equations $\dot{X}_{n}%
^{i}=-X_{n}^{i}\times(B^{i}(\mathbf{X}_{n})+b_{n}^{i})-\alpha X_{n}^{i}%
\times\lbrack X_{n}^{i}\times(B^{i}(\mathbf{X}_{n}\mathbf{)}+b_{n}^{i})],$
where $b_{n}(t)$ is a sequence of Gaussian processes which correlation
functions that go to the $\delta$-function as $n\rightarrow\infty,$ then
$\mathbf{X}_{n}$ tends to the solution $\mathbf{X}$ of (\ref{sll}) if it is
interpreted in the Stratonovich sense \cite[Chapter 2]{Strat64}, \cite[Chapter
5]{Has}). We also note in passing that one can model a Gaussian colored noise
by the Ornstein-Uhlenbeck process \cite{Gard} which can be substituted in
(\ref{sll}) instead of the white noise $b(t).$ It could be of interest to
study the influence of the correlation radius on the stochastic
Landau-Lifshitz dynamics. We do not pursue such questions in this paper but
remark that effective numerical methods for differential equations with
colored noise are available in \cite{MT,color} which can be adapted to the SLL
equation with colored noise.

Since we will exploit some results from stochastic numerics \cite{MT} which in
turn follows the standard theory of stochastic differential equations, it is
convenient to re-write the SLL equation ~(\ref{sll}) in differential form
\cite{Gard}:
\begin{eqnarray}
dX^{i}  & =X^{i}\times a_{i}(\mathbf{X})dt+X^{i}\times\sigma(X^{i})\circ
dW^{i}(t),\label{sds}\\
X^{i}(0)  & =x_{0}^{i},\ \ |x_{0}^{i}|=1,\ i=1,\ldots,n,\nonumber
\end{eqnarray}
where $W^{i}(t)=(W_{x}^{i}(t),W_{y}^{i}(t),W_{z}^{i}(t))^{\top},$
$i=1,\ldots,n;$ $W_{x}^{i}(t),\ W_{y}^{i}(t),\ W_{z}^{i}(t),$ $i=1,\ldots,n,$
are independent standard Wiener processes; $a_{i}(\mathbf{x}),$ $\mathbf{x}%
\in\mathbb{R}^{3n},$ are three-dimensional column-vectors defined by
\begin{equation}
a_{i}(\mathbf{x})=-B^{i}(\mathbf{x})-\alpha x^{i}\times B^{i}(\mathbf{x}%
)\ ;\label{sda}%
\end{equation}
and $\sigma(x),$ $x\in\mathbb{R}^{3},$ is a $3\times3$-matrix such that
\begin{equation}
\sigma(x)y=-\sqrt{2D}y-\alpha\sqrt{2D}x\times y\label{sig}%
\end{equation}
for any $y\in\mathbb{R}^{3}$. Note that the symbol $`\circ$' in Eq.~(\ref{sds}%
) means that the corresponding stochastic integral is interpreted in the
Stratonovich sense \cite{Gard}. We recall \cite{Strat64} (see also
\cite{Gard,Has}) that the Stratonovich stochastic integral can be defined as
the mean-square limit of the middle Riemann sums, which, in particular, makes
it evident why the midpoint scheme (see (\ref{imp}) below) satisfies the
Stratonovich calculus.

Let us consider some properties of the solution to (\ref{sds})-(\ref{sig}).
First, the length of each individual spin is a constant of motion, i.e.,
\begin{equation}
|X^{i}(t)|=1,\ i=1,\ldots,n,\ \ t\geq0.\label{spin_pres}%
\end{equation}
Indeed, we have
\begin{eqnarray*}
d\frac{1}{2}|X^{i}|^{2}  & =X^{i}dX^{i}\\
& =X^{i}\left[  X^{i}\times a_{i}(\mathbf{X})\right]  dt+X^{i}\left[
X^{i}\times\sigma(X^{i})\circ dW_{i}(t)\right]  =0.
\end{eqnarray*}

Other general conservation laws of (\ref{sds})-(\ref{sig}) and also of
(\ref{lln}) do not exist. However when we restrict ourselves to realistic
systems, we have the damping coefficient $\alpha\ll1$. This means that, in
practice, solutions of (\ref{sds})-(\ref{sig}) are, in a sense, close to the
deterministic solutions of (\ref{lln}) with $\alpha=0$. Hence the precessional
motion can usually be considered as dominant. In turn, the largest
contribution to the precessional motion is due to the exchange interaction.
Therefore, it is relevant to examine the conservation laws for $\alpha=0$.
Since the Hamiltonian has no explicit time-dependence, energy is conserved for
this case. Further, when only Heisenberg exchange is included we have for the
total spin:
\begin{equation}
\sum_{i}\frac{d{X}^{i}}{dt}=\sum_{i\neq j}J_{ij}\,X^{i}\times X^{j}=\sum
_{i>j}J_{ij}\,(X^{i}\times X^{j}+X^{j}\times X^{i})=0\label{totspin}%
\end{equation}
since $J_{ij}=J_{ji}$. We recall that the orientation of individual spins is
time dependent, which makes the effective field acting on each spin time
dependent due to the exchange interaction. However, at the same time, the
symmetry of the exchange interaction ensures that the total spin is
time-independent. Therefore the conservation of total spin is an important
property for stable numerical integration of the exchange interaction. By the
same arguments, when an external field is added, the total spin will precess
in the external field:
\begin{equation}
\sum_{i}\frac{d{X}^{i}}{dt}={B_{0}}\times\sum_{i}{X}^{i}.
\end{equation}
For this case, the length of the total spin is a constant of motion, as well
as the component of the total spin along $B_{0}$. Hence the energy is also
conserved but the transversal components of the total spin with respect to
$B_{0}$ oscillate in time. When anisotropy is included, there are no
conservation properties associated with the total spin. Finally, ergodicity of
the solution to (\ref{sds})-(\ref{sig}) is a relevant property. This is
discussed in Appendix~A.

\section{Numerical methods}

\label{sec_numermethod}

In this section we consider numerical integrators for the stochastic
Landau-Lifshitz equation (\ref{sds})-(\ref{sig}). We first recall two existing
numerical methods, one of which is explicit (the projected Heun scheme) and
the other implicit (the midpoint scheme). Both are unsatisfactory since either
they violate conservation laws (HeunP) or they are computationally very
expensive (IMP). Therefore, in the main part of this section we present the
two newly developed numerical methods (SIA and SIB). These methods are called
semi-implicit and aim at combining the advantages of the existing explicit and
implicit schemes.

As it is known from the deterministic (\cite{HLW02,frank1997,arponen2004} and
the references therein) and stochastic (\cite{MRT02b,MT,RRM09}) numerical
approaches, to achieve accuracy in long-time simulations (e.g., for computing
ergodic limits) it is advisable to preserve the structural properties of the
continuous dynamics by the approximating discrete ones. Then it is important
to consider not only orders of convergence but also structural properties of
numerical integrators for the SSL equation. Both convergence and structural
properties of the schemes presented are discussed in Section~\ref{secdis}.

Throughout we use (for simplicity) a uniform discretization of a time interval
$[0,t_{\star}]$ with step size $h=t_{\star}/N$. The value at the initial step
is $X_{0} ^{i}=x_{0}^{i},$ $i=1,\ldots,n$, and $X_{k}^{i}$, $i=1,\ldots,n$,
denotes the approximate solution $X^{i}(t_{k}),$ $i=1,\ldots,n$, to the SLL
equation at time $t_{k},$ $k=1,\ldots,N$.

\subsection{Existing explicit and implicit numerical methods}

\subsubsection{\textquotedblleft Heun + projection (HeunP)\textquotedblright.}

The Heun method can be seen as a predictor-corrector method. Its prediction
step, which we denote by $\mathbf{\mathcal{X}}_{k}$, is the Euler
approximation. The standard Heun method should be adjusted by an additional
projection step which is needed to ensure that the length of each individual
spin remains constant. For the SLL equation~(\ref{sds})-(\ref{sig}), the HeunP
method reads
\begin{eqnarray}
\mathcal{X}_{k}^{i}  & =X_{k}^{i}+hX_{k}^{i}\times a_{i}(\mathbf{X}%
_{k})+h^{1/2}X_{k}^{i}\times\sigma(X_{k}^{i})\xi_{k+1}^{i},\label{heunp}\\
& i=1,\ldots,n,\nonumber\\
{X^{\ast}}_{k+1}^{i}  & =X_{k}^{i}+\frac{h}{2}\left[  X_{k}^{i}\times
a_{i}(\mathbf{X}_{k})+\mathcal{X}_{k}^{i}\times a_{i}(\mathbf{\mathcal{X}}%
_{k})\right]  \nonumber\\
& +\frac{h^{1/2}}{2}\left[  X_{k}^{i}\times\sigma(X_{k}^{i})\xi_{k+1}%
^{i}+\mathcal{X}_{k}^{i}\times\sigma(\mathcal{X}_{k}^{i})\xi_{k+1}^{i}\right]
,\nonumber\\
X_{k+1}^{i}  & ={X^{\ast}}_{k+1}^{i}/|{X^{\ast}}_{k+1}^{i}|,\ i=1,\ldots
,n,\nonumber\\
& k=1,\ldots,N,\nonumber
\end{eqnarray}
where $\mathbf{\mathcal{X}}_{k}=(\mathcal{X}_{k}^{1^{\top}},\ldots
,\mathcal{X}_{k}^{n^{\top}})^{\top};$ $\xi_{k+1}^{i}=\left(  \xi_{k+1}%
^{i,1},\xi_{k+1}^{i,2},\xi_{k+1}^{i,3}\right)  ^{\top};$ $\xi_{k}^{i,j},$
$j=1,2,3,$ $i=1,\ldots,n,$ $k=1,\ldots,N,$ are independent identically
distributed (i.i.d.) random variables which can be distributed, e.g., as
\begin{equation}
P(\xi_{k}^{i,j}=\pm1)=1/2\label{simdi}%
\end{equation}
or $\xi_{l}^{i,j}\sim$ $\mathcal{N}(0,1)$. This indicates that the $\xi
_{l}^{i,j}$, are i.i.d. Gaussian random variables with zero mean and unit
variance. In Eqs.~(\ref{heunp}) we explicitly added $i=1,\ldots,n$ to
emphasize that first $\mathcal{X}_{k}$ has to be calculated for all spins,
before $\mathbf{X}_{k+1}$ is computed. We come back to this point in the
numerical experiments (Section~\ref{sec_exp}).

\subsubsection{\textquotedblleft Implicit Midpoint (IMP)\textquotedblright.}

Contrary to the HeunP method, IMP (see, e.g. \cite[p. 45]{MT})) is implicit.
For the SLL equation~(\ref{sds})-(\ref{sig}), IMP reads:
\begin{eqnarray}
X_{k+1}^{i}  &  =X_{k}^{i}+h\frac{X_{k}^{i}+X_{k+1}^{i}}{2}\times a_{i}\left(
\frac{\mathbf{X}_{k}+\mathbf{X}_{k+1}}{2}\right) \label{imp}\\
&  +h^{1/2}\frac{X_{k}^{i}+X_{k+1}^{i}}{2}\times\sigma\left(  \frac{X_{k}%
^{i}+{X}_{k+1}^{i}}{2}\right)  \xi_{k+1}^{i}\ ,\ \ i=1,\ldots,n,\nonumber\\
&  k=1,\ldots,N,\nonumber
\end{eqnarray}
where $\xi_{k+1}^{i}=\left(  \xi_{k+1}^{i,1},\xi_{k+1}^{i,2},\xi_{k+1}%
^{i,3}\right)  ^{\top};$ $\xi_{k}^{i,j},$ $j=1,2,3,$ $i=1,\ldots,n,$
$k=1,\ldots,N,$ are i.i.d. random variables which can be distributed according
to, e.g., (\ref{simdi}). Alternatively, we can choose $\xi_{k}^{i,j}$ being
distributed as the $\xi_{h}$ defined below (see \cite{MRT02b,MT}). Let $\zeta$
$\sim$ $\mathcal{N}(0,1)$ be a Gaussian random variable with zero mean and
unit variance. We define
\begin{equation}
\xi_{h}=\left\{
\begin{array}
[c]{c}%
\zeta,\;|\zeta|\leq A_{h},\\
A_{h},\;\zeta>A_{h},\\
-A_{h},\;\zeta<-A_{h},
\end{array}
\right.  \label{Fin61}%
\end{equation}
where $A_{h}=\sqrt{2|\ln h|}.$ We note that if one takes $\xi_{k}^{i,j}$
$\sim$ $\mathcal{N}(0,1),$ IMP can, in general, diverge (see a counter-example
in \cite{MRT02b,MT}).

\subsection{New semi-implicit numerical methods}

Here we propose two new semi-implicit integration schemes, simply called
semi-implicit A (SIA) and semi-implict B (SIB). In the spirit of the review
\cite{banas2005}, they are called semi-implicit since they require only to
solve $n$ or, $2n$ in the case of the SIB\ scheme, linear $3\times3$ systems
at each time-step, which can be done analytically. The starting point for
derivation of the semi-implicit methods is the IMP scheme. To reduce the
degree of implicitness, we replace $\mathbf{X}_{k+1}$ in the argument of
$a_{i}$ and $\sigma$ in IMP by a predictor $\mathbf{\mathcal{X}}_{k}$. As a
consequence, resolving the implicitness at each time step is simplified (in
comparison to IMP) to solving a linear $3\times3$ system per spin that is
independent of the interactions between the spins. The difference between SIA
and SIB is the choice for $\mathbf{\mathcal{X}}_{k}$. Both semi-implicit
methods have effectively the same computational cost as explicit schemes.

\subsubsection{\textquotedblleft Semi-implicit scheme A
(SIA)\textquotedblright.}

Similar to the HeunP method, for the SIA scheme we take the Euler
approximation for the predictor $\mathbf{\mathcal{X}}_{k}$. The SIA method for
the SLL equation reads
\begin{eqnarray}
\mathcal{X}_{k}^{i}  &  =X_{k}^{i}+hX_{k}^{i}\times a_{i}(\mathbf{X}%
_{k})+h^{1/2}X_{k}^{i}\times\sigma(X_{k}^{i})\xi_{k+1}^{i},\label{mesi}\\
& i =1,\ldots,n,\nonumber\\
X_{k+1}^{i}  &  =X_{k}^{i}+h\frac{X_{k}^{i}+X_{k+1}^{i}}{2}\times a_{i}\left(
\frac{\mathbf{X}_{k}+\mathbf{\mathcal{X}}_{k}}{2}\right) \nonumber\\
&  +h^{1/2}\frac{X_{k}^{i}+X_{k+1}^{i}}{2}\times\sigma\left(  \frac{X_{k}%
^{i}+\mathcal{X}_{k}^{i}}{2}\right)  \xi_{k+1}^{i},\ i=1,\ldots,n,\nonumber\\
&  k=1,\ldots,N,\nonumber
\end{eqnarray}
where $\xi_{k+1}^{i}=\left(  \xi_{k+1}^{i,1},\xi_{k+1}^{i,2},\xi_{k+1}%
^{i,3}\right)  ^{\top};$ $\xi_{l}^{i,j}$ are i.i.d. random variables as in
IMP~(\ref{imp}) (the same two possibilities).

\subsubsection{\textquotedblleft Semi-implicit scheme B
(SIB)\textquotedblright.}

SIA can be viewed as a second iteration for the implicit equation due to IMP.
As zero approximation of $\mathbf{X}_{k+1},$ we took $\mathbf{X}_{k}$ and then
the second iteration was constructed so that the length of individual spins is
preserved. One can see that the first iteration (or in other words the
prediction step) of SIA does not preserve the spin length. We are therefore
proposing the SIB method which keeps the spin-length conserving IMP structure
at both iterations and, according to our numerical tests (see
Section~\ref{sec_exp}), this modification is crucial for the performance of
the semi-implicit schemes.

The SIB method for the SLL equation reads%
\begin{eqnarray}
\mathcal{X}_{k}^{i}  &  =X_{k}^{i}+h\frac{X_{k}^{i}+\mathcal{X}_{k}^{i}}%
{2}\times a_{i}(\mathbf{X}_{k})+h^{1/2}\frac{X_{k}^{i}+\mathcal{X}_{k}^{i}}%
{2}\times\sigma(X_{k}^{i})\xi_{k+1}^{i},\label{ssi}\\
& i =1,\ldots,n,\nonumber\\
X_{k+1}^{i}  &  =X_{k}^{i}+h\frac{X_{k}^{i}+X_{k+1}^{i}}{2}\times a_{i}\left(
\frac{\mathbf{X}_{k}+\mathbf{\mathcal{X}}_{k}}{2}\right) \nonumber\\
&  +h^{1/2}\frac{X_{k}^{i}+X_{k+1}^{i}}{2}\times\sigma\left(  \frac{X_{k}%
^{i}+\mathcal{X}_{k}^{i}}{2}\right)  \xi_{k+1}^{i}\ ,\ i=1,\ldots
,n,\ \nonumber\\
&  k=1,\ldots,N,\ \nonumber
\end{eqnarray}
where $\xi_{k+1}^{i}=\left(  \xi_{k+1}^{i,1},\xi_{k+1}^{i,2},\xi_{k+1}%
^{i,3}\right)  ^{\top};$ $\xi_{l}^{i,j}$ are i.i.d. random variables as in
IMP~(\ref{imp}) (the same two possibilities).

\begin{remark}
One can continue the process and make\ several iterations for the implicit
equation due to IMP, e.g., in our tests about $10$ iterations were sufficient
to resolve the implicitness up to the machine accuracy. However, in practice
the use of several iterations would be too computationally expensive while SIB
already demonstrates stability and accuracy comparable with IMP.
\end{remark}

\subsection{Properties of the methods\label{secdis}}

We start by examining convergence of the methods presented in this section and
then discuss some conservation properties. For completeness, in Appendix~B we
recall some generic facts about stochastic numerics \cite{MT}.

All four methods considered in this section are of weak order one for both
choices of the distributions of $\xi_{k}^{i,j}$ (discrete and continuous). If
$\xi_{l}^{i,j}$ $\sim$ $\mathcal{N}(0,1)$, then HeunP is also of mean-square
order $1/2$. IMP, SIA, and SIB are of mean-square order $1/2$ if $\xi
_{k}^{i,j}$ have the cut-off Gaussian distribution (\ref{Fin61}). These
convergence properties are proved using the standard results \cite[Chapters 1
and 2]{MT}. In the deterministic case (i.e., $D=0$) all four methods are of
order two.

Note that in this paper we limit ourselves to methods of weak order $1$ and of
mean-square order $1/2.$ The system (\ref{sds})-(\ref{sig}) has noncommutative
noise (see the definition in, e.g. \cite[p. 28]{MT}). Then mean-square methods
of orders higher than $1/2$ require simulation of multiple Ito integrals which
is computationally expensive. It is possible to construct higher order weak
methods for (\ref{sds})-(\ref{sig}) but, due to the multiplicative,
noncommutative nature of the noise, they would be too complicated and they are
not considered here. We also note that the problem with multiplicative noise
can be circumvented by rewriting the SLL equation in spherical coordinates,
for which the system is Hamiltonian and the noise becomes additive, but then
numerical difficulties arise when the polar angle is close to $0$ or $\pi$.

When $\alpha$ is small, the SLL equation~(\ref{sds})-(\ref{sig}) is a system
with small multiplicative noise. In this case the weak-sense errors of all the
methods considered in this section are of order $O(h^{2}+\alpha^{2}h)$
\cite{sinum},\cite[Chapter 3]{MT}. The smallness of noise can be further
exploited to construct high accuracy but low order efficient methods following
the recipe from \cite{sinum,MT}.

We now discuss \textit{conservation properties} of the schemes. The HeunP
method (\ref{heunp}) has only one conservation property -- norm-preservation
which is due to the projection step. Heun without the projection step would
conserve the total spin but then violates norm-preservation. Omitting the
projection step also gives very poor results for the interaction with an
external magnetic field. In practice the projection step can be exploited for
error control. Energy is not conserved by HeunP when $\alpha=0$. HeunP has the
advantage of being very flexible, its implementation is independent of the
symmetry of the system and types of interactions used. The method is also fast
since integration can be done for each spin separately.

Due to the structure of IMP, the difference $X_{k+1}^{i}-X_{k}^{i}$ is always
perpendicular to $X_{k}^{i}+X_{k+1}^{i}$. Therefore $(X_{k}^{i}+X_{k+1}%
^{i})(X_{k+1}^{i}-X_{k}^{i})=0$ and hence $|X_{k+1}^{i}|^{2}=|X_{k}^{i}|^{2},$
i.e., the length of each spin is exactly preserved by IMP without any need of
projection. In the deterministic case with $\alpha=0$ and under only the
Heisenberg exchange, IMP conserves the total spin. The proof follows directly
from Eq.~(\ref{totspin}) with replacing $dX^{i}/dt$ by $X_{k+1}^{i}-X_{k}^{i}$
and $X^{i}$ by $(X_{k}^{i}+X_{k+1}^{i})/2$. The total energy conservation for
the case of $\alpha=0$ can be proven similarly. Preservation of all the main
structural properties of the SLL equation by IMP comes at a cost. Since all
spins are coupled, a system of $3n$ non-linear algebraic equations has to be
solved at each time step. This is a major limitation for application of IMP to
atomistic spin dynamics, where the number of spins is typically of order
$n=10^{6}$. Some further remarks on conservation properties of both HeunP and
IMP in the deterministic case are given in \cite{frank1997}.

The SIA method is very close to the HeunP method. However, unlike the HeunP
method SIA preserves the constraint $|X^{i}(t)|=1$ exactly, without the need
of projection. This follows directly from the observation that the norm
conservation of each spin is independent of the point at which $a_{i}$ and
$\sigma$ are evaluated. Let us now look at SIA in the deterministic case with
$\alpha=0.$ Regarding total spin, the relevant symmetry property is:
\[
\frac{X_{k}^{i}+X_{k+1}^{i}}{2}\times\frac{X_{k}^{j}+\mathcal{X}_{k}^{j}}%
{2}+\frac{X_{k}^{j}+X_{k+1}^{j}}{2}\times\frac{X_{k}^{i}+\mathcal{X}_{k}^{i}%
}{2}\neq0\ ,
\]
which is violated since the Euler approximation for $\mathcal{X}_{k}^{i}$
depends only on the orientation of the spins at the current time step
($\mathbf{X}_{k}$), but not on $\mathcal{X}_{k}^{i}$, whereas ${X}_{k+1}^{i}$
is also determined by the value ${X}_{k+1}^{i}$ itself. Owing to this
difference, for $\alpha=0$ the total spin cannot be preserved by SIA. Also,
the energy is not a conserved quantity by SIA and the scheme introduces
numerical damping. Hence SIA has the same conservation properties as HeunP,
and it is of interest to investigate whether the built-in norm conservation is
sufficient to improve stability properties.

Unlike SIA, SIB has the norm-conserving midpoint structure for both $X_{k}%
^{i}$ and $\mathcal{X}_{k}^{i}$. In the case of a two-spin deterministic
system with $\alpha=0$ we proved analytically that both energy and total spin
are conserved quantities of SIB. Hence for this system SIB has the same
conservation properties as IMP. At the same time, implementation-wise very
little additional computational efforts are required by SIB compared to HeunP
and SIA. Hence it is of interest to compare the performance of SIB with SIA,
in particular to investigate the influence of preservation of
norm-conservation and preservation of deterministic conservation laws on the
stability properties of the methods. As our numerical experiments (see the
next section) suggest, SIB outperforms SIA while SIA is only slightly better
than HeunP. This observation implies, in particular, that the built-in norm
conservation alone is not sufficient for obtaining superior numerical
integrators for ASD and preservation of other structural properties of the SLL
equation should guide one in constructing effective numerical methods. 

\section{Numerical experiments\label{sec_exp}}

In this section, we compare performance of the integrators introduced in the
previous section using two model problems. In Sec.~\ref{2spins}, we present
some results of the experiments in the deterministic case without damping
(\textit{i.e.} $\alpha=0$), to illustrate the conservation properties of the
numerical methods. In Sec.~\ref{1Dheis}, we consider the stochastic case using
the 1D Heisenberg chain as a test system. We show that the methods that
preserve the deterministic integration laws give rise to a more stable
integration for the stochastic spin dynamics.

\subsection{Two interacting spins}

\label{2spins}

In order to illustrate the conservation properties of the numerical schemes
related to the deterministic precessional motion, we choose the simple case of
two interacting spins with equal length $|{X}^{1}|=|{X}^{2}|=1$. As a result
of the exchange interaction, the spins rotate around a common axis, where the
precession frequency is given by $\omega_{J}=2J\cos\theta/2$ with the angle
$\theta$ between the spins and the Heisenberg exchange parameter $J$.

\begin{figure}[ptb]
\includegraphics[width=\columnwidth]{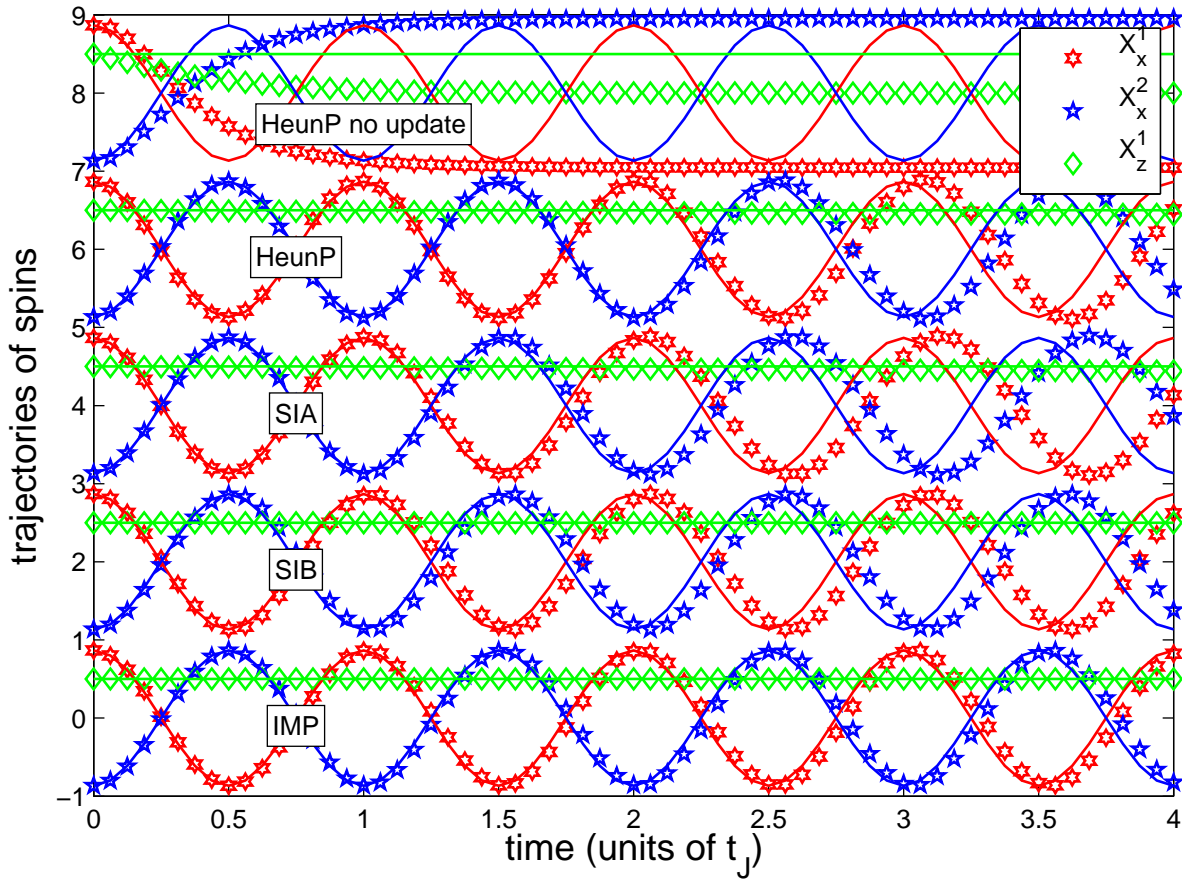} \caption{Comparison of the
explicit HeunP, implicit IMP and semi-implicit methods SIA and SIB for
the deterministic case $\alpha=0$. The trajectory of 2 interacting spins is
shown by plotting the $x$ components of the 2 spins and 1 $z$-component. 
Solid lines indicate the analytical solution.
The
upper panel shows that without simultaneous update of the effective field the
integration is very unstable. 
IMP demonstrates the best performance. All methods
introduce errors in the precession period $t_{J}=2\pi/J$ corresponding to
initial condition. For the purpose of illustration, a large step size $h=1/16$
is used.}%
\label{traj2spins}%
\end{figure}

First, we emphasize the relevance of simultaneously updating the effective
field. Due to the interaction, the effective field acting on each spin is
determined by the other spin. Therefore, when using a predictor-corrector
method like HeunP, it is highly relevant to simultaneously update the
effective fields after the prediction step before calculating the correction
step. Hence, the correction step is computed taking into account that
$a_{i}(\mathcal{X}_{k})$ depends on $\mathcal{X}_{k}^{j\neq i}$ and not on
$\mathcal{X}_{k}^{i}$ alone. Therefore, at each time step the effective field
must be computed twice. By its design, a predictor-corrector method must be
implemented in this way, otherwise it will, as a rule, become a scheme of
lower order. Figure~\ref{traj2spins} shows the computed trajectory with and
without simultaneous update for the HeunP method. To achieve a comparable
accuracy without simultaneous update of the effective field, the step size
should be decreased by a factor of $10^{2}\div10^{3}$.

In the four lowest panels of figure~\ref{traj2spins} we compare the considered
integrators implemented with simultaneous update of the effective fields. For
illustration purposes, a large step size is used ($h=1/16$). For small times,
all methods show reasonable agreement with the analytical solution, but IMP
clearly has the best performance for this system. However, even IMP, which
preserves the conservation laws instrinsically, introduces errors in the
precession frequency. Since these errors do not effect the conservation
properties of the methods, we do not consider them in detail.

\begin{figure}[ptb]
\includegraphics[width=\columnwidth]{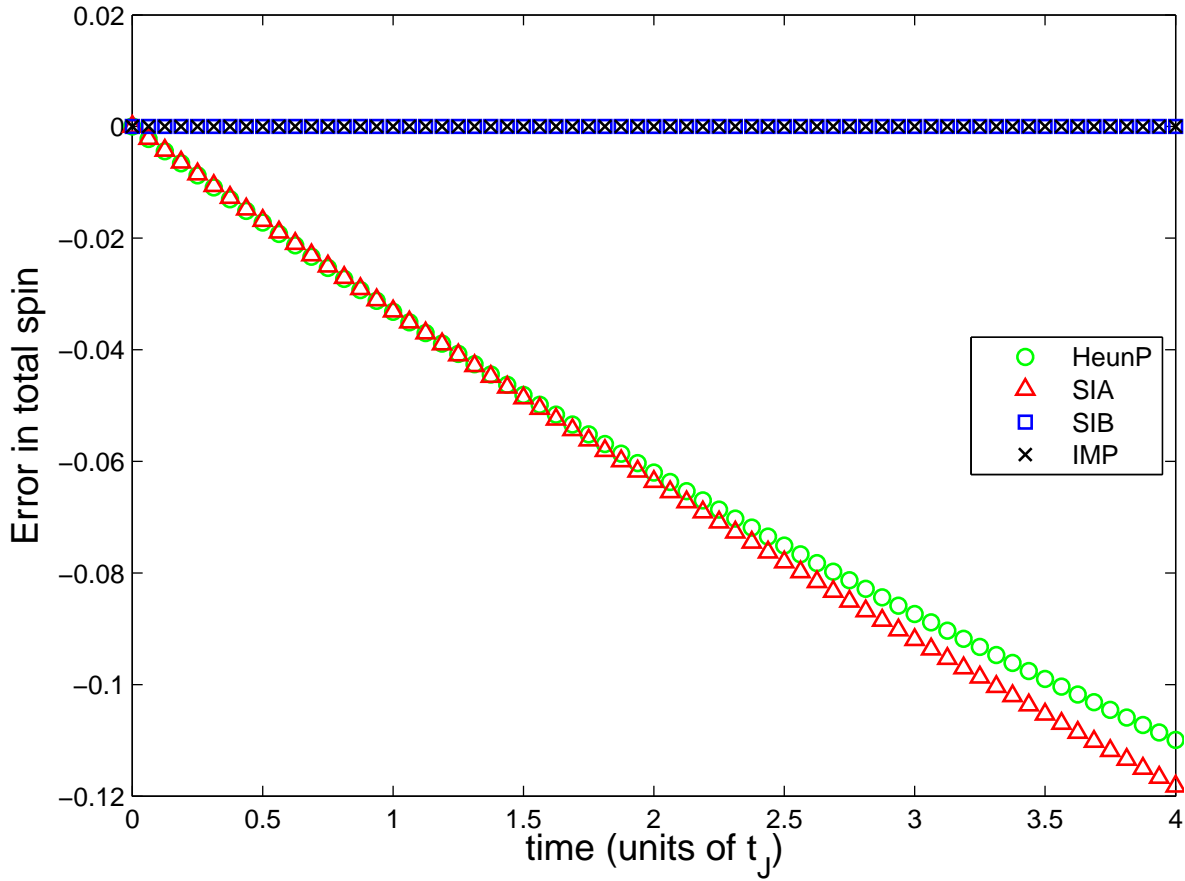}\caption{Conservation of total
spin for HeunP, IMP, SIA, and
SIB. Shown is the error in the total spin for the same system as in
figure~\ref{traj2spins}. Both IMP and SIB preserve the total spin up to
machine precision, whereas SIA and HeunP introduce a numerical damping.
Here $t_{J}=2\pi/J$ is the precession period.}%
\label{errorconservation}%
\end{figure}

Next, we compare the conservation properties of the considered methods for the
2-spin system. To this end, figure~\ref{errorconservation} shows the error in
the total spin as a function of integration time. Both SIB and IMP exactly
conserve the total spin, whereas HeunP and SIA have numerical dissipation. For
clarity, only the $z$-component of the total spin is plotted. The errors in
the $x$ and $y$-components of the total spin are much smaller since the
numerical errors in the $x,y$ motion of the individual spins cancel each other
due to the symmetry.

Despite the fact that SIA conserves the norm of each spin exactly, the
numerical damping is slightly larger than for HeunP. Both their errors are strongly
dependent on the initial condition. When the spins are almost parallel, HeunP
has a larger numerical error than SIA since the projection step transforms a
larger amount of transverse motion to longitudinal motion. In the case of
figure~\ref{errorconservation} an initial condition with $\theta
_{0}=120^{\circ}$ is used, which is closer to anti-parallel motion and,
therefore, HeunP has a smaller error than SIA.

For this simple 2-spin system, the energy and total spin are directly related:
$(X_{k}^{1}+X_{k}^{2})^{2}=(X_{k}^{1})^{2}+(X_{k}^{2})^{2}+2X_{k}^{1}X_{k}%
^{2}=2+E_{k}/J$. Hence both SIB and IMP conserve energy, whereas both HeunP
and SIA dissipate energy. For larger systems with only nearest neighbor interactions, 
SIB conserves total spin and energy like IMP as well, while obviously SIB requires
much lower computational efforts than IMP. The conservation properties of SIB can be 
proven analytically but this is beyond the scope of the present paper.

In conclusion, the results of the numerical experiments with 2 interacting
spins and $\alpha=0$ show that both HeunP and SIA introduce numerical errors
in the conserved quantities whereas SIB and IMP preserve the total spin and
energy of the test system.

\subsection{1D Heisenberg chain}

\label{1Dheis} In this section we compare the semi-implicit integration
schemes with the explicit and implicit methods in the stochastic case. The
simplest model of classical interacting spins is the 1D Heisenberg chain with
nearest-neighbor interactions. For this system, an analytical expression for
the mean energy per spin is available \cite{shubin1936,fisher1964}:
\begin{equation}
\overline{H}_{\mathrm{analytic}}\equiv\frac{\left\langle H_{\mathrm{ex}%
}\right\rangle }{2nJ}=\left(  1-\frac{1}{n}\right)  \left(  \frac{k_{b}T}%
{2J}-\coth\left(  \frac{2J}{k_{b}T}\right)  \right)  .\label{heisenergy}%
\end{equation}
This expression gives us a convenient way to check how accurately the
temperature of the system is reproduced in simulations using the numerical
methods from Sec.~\ref{sec_numermethod}. Note that $\overline{H}%
\rightarrow-1+1/n$ as the temperature $T\rightarrow0$ since we have normalized
the energy with the number of spins $n$ and the interaction energy of 2 spins
$2JX^{1}X^{2}$ tends to $2J$ when the temperature goes to zero.

\begin{figure}[ptb]
\includegraphics[width=\columnwidth]{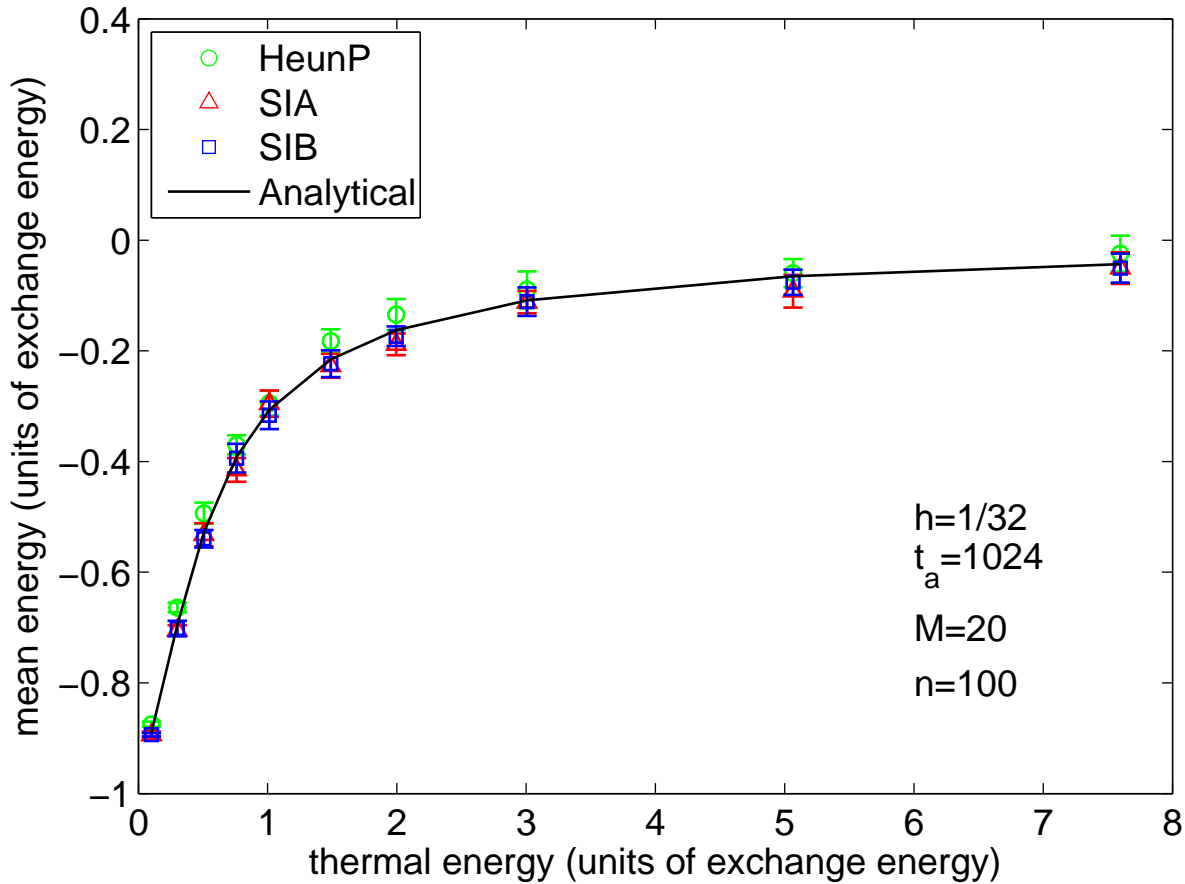}\caption{Temperature check of of
the semi-implicit methods SIA and SIB compared with the explicit HeunP method.
Shown is the mean energy per spin of the 1D Heisenberg chain, as function of
temperature, computed with the parameter values shown at the bottom. All the
schemes demonstrate reasonable agreement with the analytical result (\ref{heisenergy}%
).}%
\label{energyheis1D}%
\end{figure}

The comparison of the HeunP method with the semi-implicit schemes for the
temperature is shown in figure~\ref{energyheis1D} for step size $h=1/32$,
damping $\alpha=0.1$, exchange parameter $J=1$, spin length $|X^{i}|=1$, and
number of spins $n=100$. The random variables used in the numerical schemes
are simulated according to the cut-off Gaussian distribution (\ref{Fin61}). At
a time step $k$ the sample average $\hat{H}_{k}$ for the energy $H$ per spin
is computed as
\begin{equation}
\hat{H}_{k}=\frac{1}{M}\sum_{m=1}^{M}\frac{H_{\mathrm{ex}}(\mathbf{X}%
_{k}^{(m)})}{2nJ}\ ,\label{hatH}%
\end{equation}
where $\mathbf{X}_{k}^{(m)}$ are independent realizations of $\mathbf{X}_{k}$
obtained by a numerical scheme (see also Appendix~B). The corresponding
standard deviation $\sigma_{H_{k}}$ is also computed. In the experiment an
ensemble of $M=20$ independent trajectories was used. The values plotted in
figure~\ref{energyheis1D}, with the $95\%$ confidence intervals determined by
the standard deviation, were obtained after equilibrating the system for a
time $t_{a}=1024\,t_{J}$, long enough for the system to be sufficiently close
to equilibrium. Here $\,t_{J}=2\pi/(2J)$ is the reference precession period
for (almost) parallel spins. We find that both HeunP and the semi-implicit
schemes show reasonable agreement with the analytical results, indicating that
they obey the Stratonovich interpretation rule as expected.

\begin{figure}[ptb]
\includegraphics[width=\columnwidth]{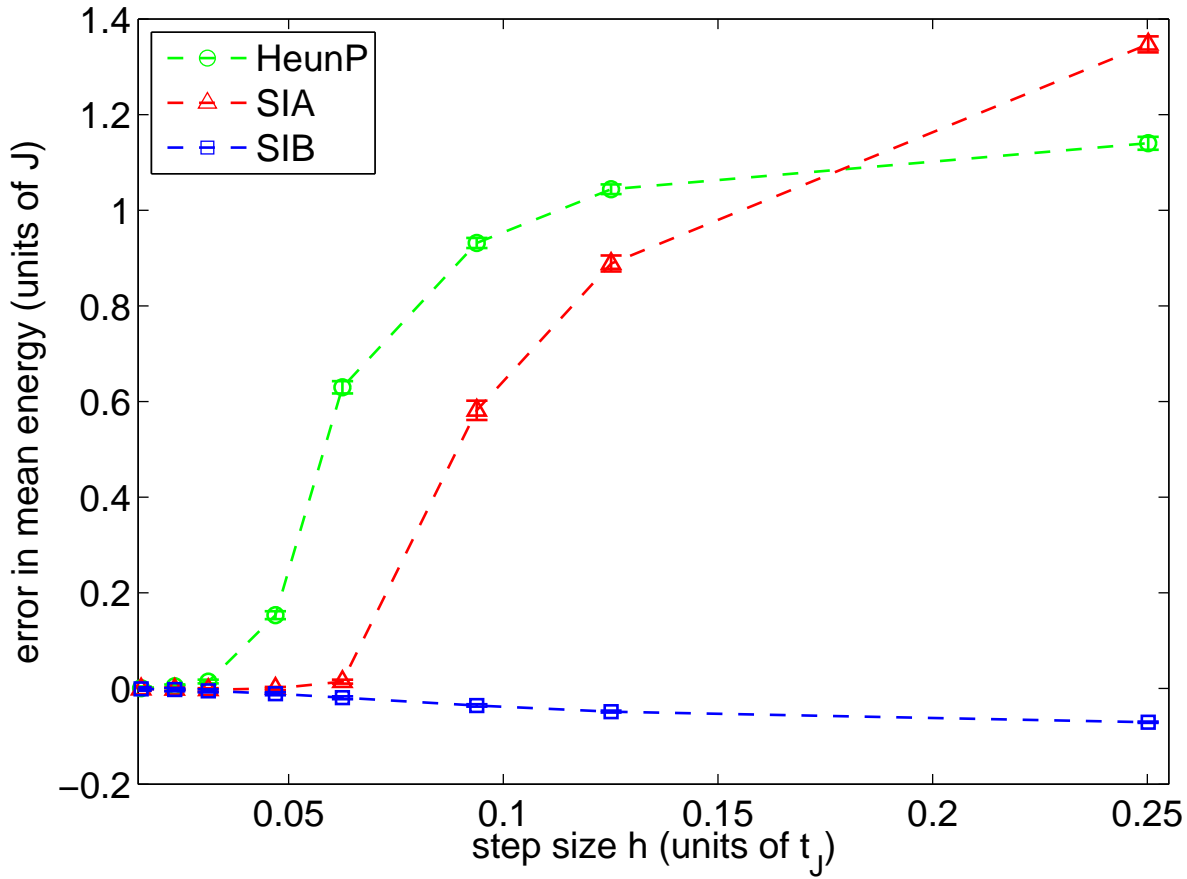}\caption{Stability of the
semi-implicit methods SIA and SIB comared with the explicit HeunP method.
Shown is the error in the mean energy as a function of the step size $h$ for
the lowest temperature considered in figure~\ref{energyheis1D}, $k_{b}%
T/(2J)=0.1$. It is found that SIB remains stable up to 4 steps per precession
period $t_{J}=2\pi/(2J)$, while SIA and HeunP become unstable and produce
unreliable results.}%
\label{convergence1}%
\end{figure}

The next question is which method is more accurate. Figure~\ref{energyheis1D}
shows that SIB is consistent with the analytical solution at all data points.
To the contrary, HeunP and SIA show slight discrepancies. To investigate this
more accurately, we study the numerical error by varying the step size. For
illustration, we used the lowest temperature $k_{b}T/(2J)=0.1$. The results
are shown in figure~\ref{convergence1}.

It is found that SIB outperforms both SIA and HeunP, and SIB remains stable down 
to only $4$ steps per precession period. At such a large step size, SIA and HeunP
are unstable though SIA performs slightly better than HeunP. Note that in
physical units, with the exchange energy $J\hat{X}^{2}=1$~mRy, $\hat{X}%
=1\mu_{\mathrm{Bohr}}$, $4$ steps per precession period corresponds to a step
size of about $20$~fs. Hence, for SIB the step size is only limited by the
precession period of the spins, and there is no need to decrease the step size
to preserve the conservation laws accurately enough. This should be compared
with the step size of 10 as which was reported in \cite{skubic2008}, resulting
in an enormous improvement of a factor $2\cdot10^{3}$ in the allowed step
size. However, the mentioned implementation of ASD in \cite{skubic2008} is
based on HeunP without the simultaneous update of the effective field. As
follows from figure~\ref{traj2spins} and figure~\ref{convergence1}, when the
effective field is properly updated, HeunP also allows a larger step size.
However, the increase is limited to about $2$ fs for the system studied here.
Compared to HeunP, SIA has only slightly better stability properties, which we
attribute to the intrinsic norm conservation. The superior stability
properties of SIB can apparently be explained by its built-in deterministic
conservation properties. For the system studied here, SIB allows  step sizes
by about a factor of $10$ larger than HeunP and by about a factor of $5$
larger than SIA.

Let us now compare the performance of the semi-implicit methods with the full
implicit IMP. The 1D Heisenberg chain is not convenient for this purpose,
unless we choose a very small number of spins. In addition, for this
comparison stability is not the major issue since we already know that the
step size of SIB is limited only by the precession period. Therefore, we are
more interested in the intrinsic properties of the integrators that are
independent of the system under study. Hence the relevant property here is the
convergence of the semi-implicit and IMP schemes. To reduce computational
costs of the experiment, we again use a system with only 2 spins.

\begin{figure}[ptb]
\includegraphics[width=\columnwidth]{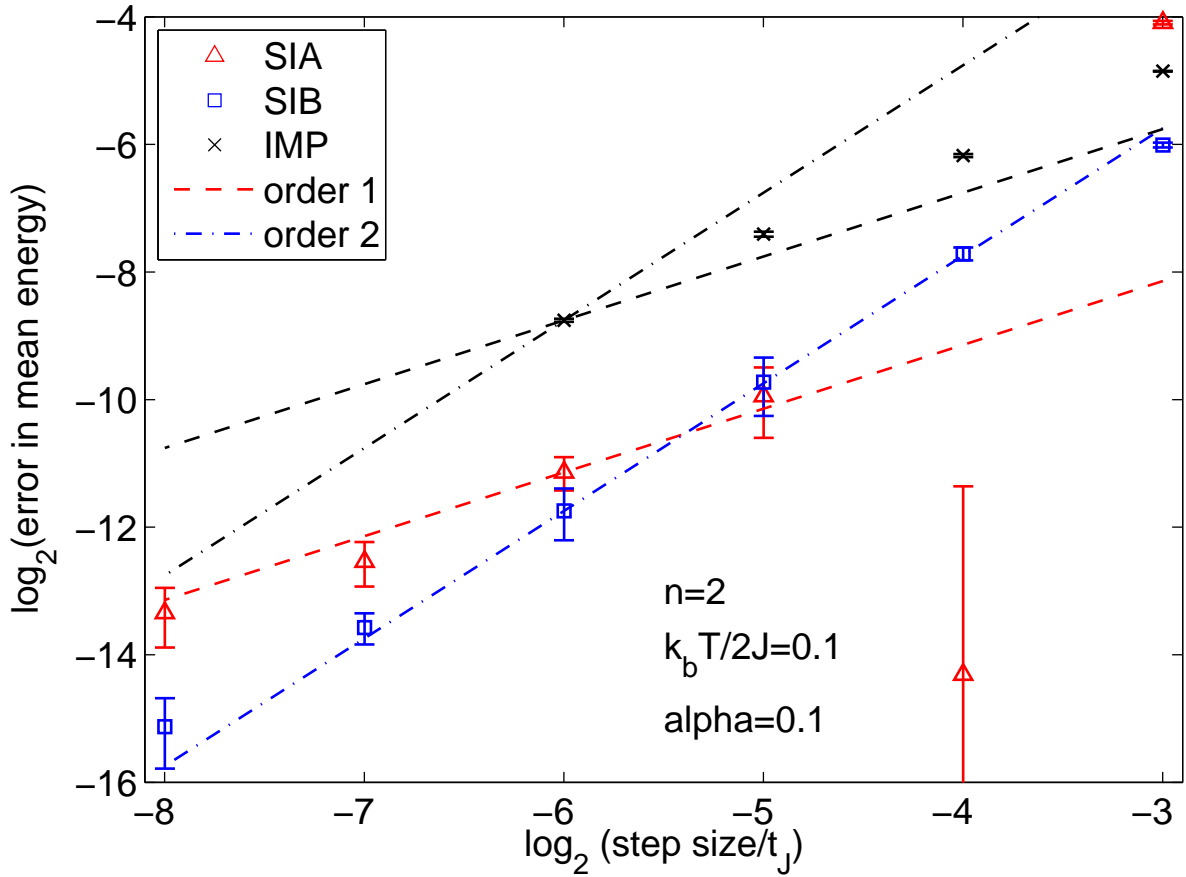}\caption{Comparison of the
semi-implicit methods SIA and SIB with the full implicit IMP. Shown is the
weak-order convergence of SIA, SIB and IMP schemes for the mean energy per
spin. Both axes are logaritmic with base 2. For small enough step size, the
slope gives the order of convergence. Surprisingly, both SIA and SIB are more
accurate than IMP. Moreover, SIB shows a higher order convergence than IMP.
Here $t_{J}=2\pi/(2J)$ indicates the reference precession period.}%
\label{convergence2}%
\end{figure}

To experimentally observe the order of convergence, a small statistical error
is needed. To this end, a combination of ensemble and time averaging was used.
As before, for an ensemble with $M$ trajectories, we let the system
equilibrate for a time $t_{a}=2048\,t_{J}$. Subsequently, the equilibrated
sample mean $\hat{H}_{k}$ (see (\ref{hatH})) is calculated for a time
$t_{b}=6144\,t_{J}$. The calculated values of $\hat{H}_{k}$ are then divided
in $P=8$ subsets of length $L=t_{b}/P=768\,t_{J}$ and in each subset the time
mean $\check{H}_{p}$ is computed. Eventually, the total mean $\check{H}$ is
the average of the time means over the $P$ subsets and its statistical error
$\Delta$ is estimated by two standard deviations of $\check{H}_{p}$ divided by
$\sqrt{P}$, which gives half of the length of the $95\%$ confidence intervals 
for $\check{H}$.
The results are presented in figure~\ref{convergence2}.

Note that for this small system no instabilities appear in SIA, and this
method shows the first weak-order convergence as expected. Surprisingly, SIB
demonstrates a second-order convergence, which might be related to the fact
that the energy is a conserved quantity for $\alpha=0$. This means that for
the energy only numerical errors from the damping term show up, hence the
convergence for the energy in the stochastic case might be better than the
convergence for a general quantity. The small error for SIA at the one but
smallest time step in figure~\ref{convergence2} is caused by the change in
sign of the error. The error values are given in Table~\ref{tab_convergence2}.
Here also the data for HeunP are provided. HeunP is not shown in
figure~\ref{convergence2} since it appears to be in the asymptotic regime only
for the smallest time steps. We note that there is a sign change of the HeunP
error, which is the reason for its small error at $h=1.564\times10^{-2}$.  IMP
is very costly for a large ensemble, therefore the two smallest step sizes
were not computed.

\begin{table}[h]\footnotesize
\caption{The values of error in the mean energy $\epsilon=\check{H}-\overline
{H}_{\mathrm{analytic}}$ and the corresponding statistical error $\Delta$
for the considered schemes. In each consecutive row the step size is smaller
by a factor 2.}%
\label{tab_convergence2}%
\raggedleft
\begin{tabular}{cccccccccc}
\br
&  & HeunP &  & SIA &  & SIB &  & IMP & \\
\mr
$h$ & $M$ & $\epsilon$ & $\Delta$ & $\epsilon$ & $\Delta$ & $\epsilon$ 
& $\Delta$ & $\epsilon$ & $\Delta$\\
1.251e-1 & $2^{4}$ & 4.23e-1 & 1.1e-3 & 5.87e-2 & 1.1e-3 & -1.55e-2 & 4.1e-4 &
-3.46e-2 & 1.3e-4\\
6.255e-2 & $2^{5}$ & 2.71e-2 & 6.0e-4 & -4.92e-5 & 3.3e-4 & -4.76e-3 &
3.3e-4 & -1.38e-2 & 2.3e-4\\
3.128e-2 & $2^{7}$ & 1.84e-3 & 4.5e-4 & -1.02e-3 & 3.7e-4 & -1.18e-3 &
3.6e-4 & -5.77e-3 & 1.4e-4\\
1.564e-2 & $2^{12}$ & -9.74e-6 & 8.1e-5 & -4.43e-4 & 7.9e-5 & -2.92e-4 &
5.9e-5 & -2.31e-3 & 4.0e-5\\
7.819e-3 & $2^{16}$ & -9.55e-5 & 3.9e-5 & -1.68e-4 & 4.0e-5 & -8.20e-5 &
1.4e-5 &  & \\
3.909e-3 & $2^{17}$ & -8.24e-5 & 3.0e-5 & -9.62e-5 & 3.0e-5 & -2.79e-5 &
1.1e-5 &  & \\
\br
\end{tabular}
\end{table}

In general, the performance of SIB in the experiments has been better than
SIA. Interestingly, despite the excellent stability of IMP, the accuracy of
IMP in the stochastic case lags behind SIB and SIA. This is a good example of
a situation when a method with better stability not necessarily has a better
accuracy. It was also observed in the deterministic case with damping that SIB
sometimes shows better accuracy than IMP. This implies that in the case of
damped motion the numerical integration error of IMP can be larger than for
SIB, as it is observed in the stochastic case. These results show that at
least for the systems considered here, SIB has the same stability properties
as IMP, but at considerable lower computational costs.

In conclusion, we find that in the stochastic case the semi-implicit method B,
with built-in deterministic conservation laws is more stable and has smaller
numerical errors than both the SIA and the HeunP method. Surprisingly, in the
stochastic case SIB is even better than IMP in terms of accuracy and
convergence. SIA performs only slightly better than HeunP in the stochastic
case, and from this we find that norm-conservation is not the most important
criterion for stable numerical integration of the SLL equation. Hence, SIB
combines the advantages of both HeunP and IMP, being both fast and stable as
well as universal. For systems with only nearest neighbor interactions, SIB
allows step sizes by a factor of 10 larger than the popular HeunP scheme, and a 
factor of $2\cdot10^3$ larger than the HeunP method without simultaneous update of the effective field.
Since in practice nearest-neighbor interactions dominate, SIB is
expected to be also advantageous for systems with more than nearest-neighbor interactions.

\section{Conclusions and Outlook}

In this paper we introduced two new semi-implicit integrators (SIA and SIB)
for stochastic Atomistic Spin Dynamics (ASD) simulations. These schemes
combine the advantages of the standard explicit projected Heun method (HeunP)
and the fully implicit midpoint method (IMP). The semi-implicit methods are
fast as explicit schemes since they require only the solution of $3$ linear
coupled equations for each spin individually and therefore they are
effectively explicit. For stability, the most important conservation law is
apparently the preservation of the total spin for the case without damping.
Like IMP, SIB preserves this conservation law for the dominant interactions in
the system and the stability properties of SIB are comparable with IMP. SIA,
which has norm-conservation built-in but not the deterministic conservation
laws, shows only slightly better stability than HeunP in the stochastic case.
Therefore, we recommend the use of SIB for ASD simulations.

Owing to the enhanced stability, larger step sizes can be used with SIB. From
our numerical experiments we can conclude that the step size can be increased
by a factor of about $10$ compared to the explicit HeunP. For SIB, the step
size is only limited by the precession frequency of the individual atomic
spins in the exchange field, which allows for step sizes of about one fourth
of the precession period which can be as large as $20$~fs. This value of the
step size has to be compared with the $10$~as that was reported for a standard
implementation of ASD simulations \cite{skubic2008}, which is based on the
HeunP method without the simultaneous update of the effective field. Hence,
the factor $2\cdot10^{3}$ improvement can be attributed to a proper update of
the effective field and built-in conservation of the total spin for SIB.
Interestingly, numerical experiments indicate that SIB can also be more
accurate than IMP in the stochastic case. Further checks for the stochastic
case, including larger systems, more complicated interactions, and
correlations, will be discussed in a following paper.

Future work should study the conservation properties of SIB in more detail in
order to give a further explanation of its excellent behavior. It would also
be of interest to obtain a method obeying conservation laws for systems with
more complicated interactions (\textit{e.g.} next-nearest neighbor,
anisotropy). In addition, one might exploit the fact that the damping motion
and the precessional motion are always perpendicular, which potentially can be
used to design an integrator that exactly dissipates energy like in continuous
dynamics. Another direction which we can pursue in future is to derive
stochastic counterparts of the geometric integrators proposed in
\cite{frank1997,arponen2004} for deterministic Landau-Lifshitz equations.
Though they lack flexibility to deal with models with arbitrary lattice
structures, such geometric integrators are expected to be highly efficient
when it is sufficient to include only nearest neighbor interaction in the
stochastic model. Our method can also be of value for micromagnetic
simulations and we expect that similar techniques can be exploited for other
physical systems, where interactions between particles are governed by a
global conservation law, e.g., systems based on diffusion equations such as
the Schr\"{o}dinger equation.

\ack This work was partially supported by the Nederlandse Organisatie voor
Wetenschappelijk Onderzoek (NWO) and de Stichting voor Fundamenteel Onderzoek
der Materie (FOM). The authors thank O. Eriksson, Uppsala University, Sweden,
for computational support.

\appendix


\section{}

In this Appendix we discuss the ergodicity of the solution $\mathbf{X}(t)$ to
(\ref{sds})-(\ref{sig}). For the solution $\mathbf{X}(t)$ of (\ref{sds}%
)-(\ref{sig}), we will also use the notation $\mathbf{X}_{\mathbf{x}}(t)$ to
reflect the dependence on the initial condition $\mathbf{X}_{\mathbf{x}%
}(0)=\mathbf{x.}$ Taking into account (\ref{ham}) and (\ref{Ham2}), we observe
that the coefficients of (\ref{sds})-(\ref{sig}) are smooth functions and due
to (\ref{spin_pres}) they remain bounded for all $t\geq0$.

One can show \cite{Has,Soize} that for $D>0$ and $\alpha>0$ the process
$\mathbf{X}(t)$ is ergodic, i.e., there exists a unique invariant measure
$\mu$ of $\mathbf{X}$ and independently of $\mathbf{x}\in\mathbb{R}^{3n}$
there exists the limit%
\begin{equation}
\lim_{t\rightarrow\infty}\left\langle \varphi(\mathbf{X}_{\mathbf{x}%
}(t))\right\rangle =\int\varphi(\mathbf{x})\,d\mu(\mathbf{x})\equiv
\varphi^{erg} \label{PA31}%
\end{equation}
for any function $\varphi(x)$ with polynomial growth at infinity. Indeed, the
solution $\mathbf{X}(t)$ of (\ref{sds})-(\ref{sig}) lives on the compact due
to (\ref{spin_pres}). Then to prove ergodicity, it is enough to show that
there is sufficient mixing. When $\alpha=0,$ the stochastic perturbation is
only precessional and, in general (e.g., for constant $B)$ the process
$\mathbf{X}(t)$ is not ergodic. When $\alpha>0,$ the stochastic perturbation
acts in all the directions on the spheres $|x^{i}|=1$ and so ensures a mixing
sufficient for the ergodicity.

We also recall the ergodic theorem, which gives the equivalence between the
ensemble and time averaging:
\begin{equation}
\lim_{t\rightarrow\infty}\frac{1}{t}\int\limits_{0}^{t}\varphi(\mathbf{X}%
_{\mathbf{x}}(s))ds=\varphi^{erg}\; \mathit{\; almost \; surely,} \label{PB51}%
\end{equation}
where the limit does not depend on $\mathbf{x}.$

Further, the invariant measure associated with the solution $\mathbf{X}(t)$ of
(\ref{sds})-(\ref{sig}) is Gibbsian with the density
\begin{equation}
\rho(\mathbf{x})\varpropto\exp(-\beta H(\mathbf{x}))\ ,\label{invden}%
\end{equation}
where $\beta=\hat{X}\hat{B}/(k_{B}T)>0$ is the inverse temperature if we
choose the noise intensity $D$ as in (\ref{dd}). To check that (\ref{invden})
is the density of the invariant measure for (\ref{sds})-(\ref{sig}) and
(\ref{dd}), one needs to verify that this $\rho(\mathbf{x})$ is the solution
of the stationary Fokker-Planck equation for (\ref{sds})-(\ref{sig}),
(\ref{dd}). Such calculations are available, e.g. in \cite{Gar98}.


\section{}

In this Appendix we recall some generic facts from stochastic numerics
\cite{MT}. In particular, we define the weak order of convergence of numerical
methods for stochastic differential equations (SDEs) and discuss errors
arising in computing ergodic limits.

Let us introduce a system of SDEs of a general form
\begin{equation}
dX=\alpha(X)dt+\sum_{l=1}^{r}\beta_{l}(X)dW_{l}(t),\ X(0)=x, \label{PA1}%
\end{equation}
where $X,$\ $\alpha,$\ $\beta_{l}$ are $d$-dimensional column-vectors and
$W_{l}(t),\ l=1,\ldots,r,$ are independent standard Wiener processes. Consider
a numerical method for (\ref{PA1}) based on the one-step approximation:%
\begin{equation}
X_{t,x}(t+h)\simeq\bar{X}_{t,x}(t+h)=x+A(t,x,h;\xi),\ 0\leq t<t+h\leq
t_{\star}, \label{PB21}%
\end{equation}
where $\xi$ is a random vector with moments of a sufficiently high order and
$A$ is a $d$-dimensional vector function. Introduce (for simplicity) the
equidistant partition of the time interval $[0,t_{\star}]$ into $N$ parts with
the step $h=t_{\star}/N$: $0=t_{0}<t_{1}<\cdots<t_{N}=t_{\star}$,
$\ t_{k+1}-t_{k}=h.$ According to (\ref{PB21}), we construct the sequence
\begin{equation}
X_{0}=x,\ X_{k+1}=X_{k}+A(t_{k},X_{k},h;\xi_{k+1}),\ k=0,\ldots,N-1,
\label{PB22}%
\end{equation}
where $\xi_{1}$ is independent of $X_{0}$ and $\xi_{k+1}$ for $k>0$ is
independent of $X_{0},\ldots,X_{k}$, $\xi_{1},\ldots,\xi_{k}.$

We note that (\ref{PB22}) contains both explicit and implicit one-step
schemes. In explicit integration schemes the approximate solution at the next
time-step, $X_{k+1}$, can be computed explicitly from the previous time-step
value $X_{k}$. For implicit methods, $A(t,x,h;\xi)$ is a solution of an
implicit relation with respect to $x,$ i.e., implicit schemes in general
require additional work.

We usually distinguish two types of convergence of numerical methods for SDEs:
mean-square (also called strong) and weak \cite{MT}. Mean-square methods are
used for direct simulation of SDEs' trajectories which, e.g., can give
information on general behavior of a stochastic model. Weak methods are
sufficient for evaluation of mean values and are simpler than mean-square
ones. We say that the method (\ref{PB22}) is weakly convergent with order
$p>0$ if
\begin{equation}
|\left\langle \varphi(X_{N})\right\rangle -\left\langle \varphi(X(t_{\star
}))\right\rangle |\leq Ch^{p} \label{A20}%
\end{equation}
for functions $\varphi$ which, together with their derivatives of a
sufficiently high order, have growth at infinity not faster than polynomial.
If a method converges with an order $p$ in the mean-square sense, it also
converges in the weak sense with order equal to or larger than $p.$ The
opposite is not true. Since weak methods suffice for computing averages, they
are appropriate for the purposes of this paper.

To evaluate the expectation $\left\langle \varphi(X_{N})\right\rangle $ on a
computer, one can apply the Monte Carlo technique:
\begin{equation}
u\equiv\left\langle \varphi(X(t_{\star}))\right\rangle \simeq\bar{u}%
\equiv\left\langle \varphi(X_{N})\right\rangle \simeq\hat{u}\equiv\frac{1}%
{M}\sum_{m=1}^{M}\varphi(X^{(m)}_{N})\ , \label{i16}%
\end{equation}
where $X^{(m)}_{N},$ $m=1,\ldots,M,$ are independent realizations of the
random variable $X_{N}.$ In (\ref{i16}) the first approximate equality
involves the numerical integration error (cf. (\ref{A20})) and the error in
the second approximate equality (the statistical error) comes from the Monte
Carlo technique.

The error of the Monte Carlo method in (\ref{i16}) is evaluated by%
\[
\bar{\Delta}=c\,\frac{\left[ \mathrm{{Var}\left\{ \varphi(X_{N})\right\}
}\right]  ^{1/2}}{M^{1/2}}\ ,
\]
where, e.g., the values $c=1,2,3$ correspond to the fiducial probabilities
$0.68,$\ $0.95,$\ $0.997,$ respectively, with the practical implication that
\begin{eqnarray}
\bar{u}  &  \in(\hat{u}-\frac{c}{\sqrt{M}}\sqrt{\hat{v}},\hat{u}+\frac
{c}{\sqrt{M}}\sqrt{\hat{v}})\ ,\label{i20}\\
\hat{v}  &  \equiv\frac{1}{M}\sum_{m=1}^{M}\left[  \varphi(_{m}X_{N})\right]
^{2}-\hat{u}^{2}\ ,\nonumber
\end{eqnarray}
with probability $0.68$ for $c=1,$ $0.95$ for $c=2,$ and $0.997$ for $c=3.$

Now we assume that the solution of (\ref{PA1}) is ergodic. In computing
ergodic limits an additional error arises. We note that ergodic limits can be
computed using the ensemble averaging or time averaging. In the former case it
follows from a relation of the form (\ref{PA31}) for the solution $X(t)$ of
(\ref{PA1}) that for any $\varepsilon>0$ there exists $t_{a}>0$ such that for
all $t_{\star}\geq t_{a}$
\begin{equation}
\left\vert \left\langle \varphi(X_{x}(t_{\star}))\right\rangle -\varphi
^{erg}\right\vert \leq\varepsilon. \label{PA35}%
\end{equation}
Then we can use the following estimator for the ergodic limit $\varphi^{erg}%
$:
\begin{equation}
\varphi^{erg}\approx\left\langle \varphi(X_{x}(t_{\star}))\right\rangle
\approx\left\langle \varphi(X_{N})\right\rangle \approx\hat{\varphi}%
^{erg}\equiv\frac{1}{M}\sum_{m=1}^{M}\varphi\left(  X^{(m)}_{N}\right)  \ ,
\label{S1}%
\end{equation}
where the first approximate equality corresponds to the time cut-off while the
second one relates to the numerical integration error, and the third to the
statistical error as before. In this ensemble-averaging approach each of the
errors is controlled by its own parameter (see \cite{MT7}).

The time-averaging approach to computing ergodic limits is based on a relation
of the form (\ref{PB51}). By approximating a single trajectory, one gets for a
sufficiently large $\tilde{t}_{\star}$:%
\begin{equation}
\varphi^{erg}\sim\frac{1}{\tilde{t}_{\star}}\int\limits_{0}^{\tilde{t}_{\star
}}\varphi(X_{x}(s))ds\sim\check{\varphi}^{erg}\equiv\frac{1}{L}\sum_{k=1}%
^{L}\varphi(X_{k}),\label{PB52}%
\end{equation}
where $Lh=\tilde{t}_{\star}.$ Let us emphasize that $\tilde{t}_{\star}$ in
(\ref{PB52}) is much larger than ${t}_{\star}$ in (\ref{S1}) because
$\tilde{t}_{\star}$ should be such that it not just ensures the distribution
of $X(t)$ to be close to the invariant distribution (like it is required from
${t}_{\star}$) but it should also guarantee smallness of the variance of
$\check{\varphi}^{erg}$. See further details about computing ergodic limits
in, e.g. \cite{MT7,JAM,Tal90} and the references therein.

\section*{References}

\bibliographystyle{unsrt}
\bibliography{references-jpcm}

\end{document}